\documentclass[fleqn,twoside]{article}
\usepackage{espcrc2}
\usepackage{graphicx}
\usepackage[figuresright]{rotating}

\def\ra{\rightarrow}

\newcommand{\num}{\nu_{\mu}}
\newcommand{\nut}{\nu_{\tau}}

\title{MACRO constraints on violation of Lorentz invariance}

\author{M. Cozzi\address{Dipartimento di Fisica dell'Universit\`a di Bologna and INFN, 40127 Bologna, Italy}}
        
\begin{document}

\begin{abstract}

The energy spectrum of neutrino-induced upward-going muons in MACRO
has been analysed
in terms of relativity principles violating effects, keeping standard
mass-induced atmospheric neutrino oscillations as the dominant source of
$\nu_{\mu} \rightarrow \nu_{\tau}$ transitions. The data disfavor these exotic
possibilities even at a sub-dominant level, and stringent 90\% C.L.
limits are placed on the Lorentz invariance violation parameter
$|\Delta v| < 6 \times 10^{-24}$ at $\sin 2{\theta}_v$ = 0 and
$|\Delta v| < 2.5 \div 5 \times 10^{-26}$ at $\sin 2{\theta}_v$ = $\pm$1.
These limits can also be re-interpreted as upper bounds on the
parameters describing violation of the Equivalence Principle.
\end{abstract}

\maketitle

\section{Introduction}

Neutrino mass-induced oscillations are the best explanation of the atmospheric neutrino 
problem \cite{macro-98,macro-last,sk-dip,soudan2}. Two flavor $\num \ra \nut$
oscillations are strongly favored over a wide range of alternative solutions \cite{macro-sterile,sk-sterile,sk-habig,vlad-oujda,fogli,glashow9799,eclipse}. 
%
%
These alternative mechanisms have been considered under the hypothesis
that each one of them solely accounts for the observed effects. 
We address the possibility of a mixed scenario: one mechanism,
the mass-induced flavor oscillations, is considered dominant and a second
mechanism is included in competition with the former. We studied, as sub-dominant
mechanism, neutrino flavor transitions induced by violations of relativity principles,
i.e. violation of the Lorentz invariance (VLI) or of the equivalence principle (VEP).

In this mixed scenario, we assume that neutrinos can be described in terms of three
distinct bases: flavor eigenstates, mass eigenstates and velocity eigenstates,
the latter being characterized by different maximum attainable velocities (MAVs), and
consider that only two families contribute to the atmospheric neutrino oscillations.
When both mass-induced and VLI-induced neutrino oscillations are considered
simultaneously, the $\nu_\mu$ survival probability can be
expressed as~\cite{fogli,glashow9799}
\begin{equation}
P_{\nu_\mu \to \nu_\mu} = 1 - \sin^2 2 \Theta \sin^2 \Omega
\label{eq:due}
\end{equation}
where $\Theta$ and  $\Omega$ are given by:
\begin{equation}
2\Theta = \arctan (a_1/a_2); \hspace{0.2 cm} \Omega = \sqrt{ a_1^2 + a_2^2}~.
\label{eq:tre}
\end{equation}
The terms $a_1$ and $a_2$ contain the relevant physical information and the whole domain of variability of the parameters can be accessed with the requirements
$\Delta m^{2} \ge 0$, $0 \le \theta_m \le \pi/2$, $\Delta v \ge 0$
and $-\pi/4 \le \theta_v \le \pi/4$.
The functional form of the
oscillation probabilities for mass induced (VLI induced) oscillations exhibits an $L/E_{\nu}$ ($L E_{\nu}$) dependence on the
neutrino energies and pathlengths respectively.
The same formalism also applies to violation of the equivalence principle,
after substituting $\Delta v/2$ with the adimensional product $|\phi| \Delta \gamma$;
$\Delta \gamma$ is the difference of the coupling constants for neutrinos
of different types to the gravitational potential $\phi$~\cite{gasperini}.

\begin{figure}[htb]
\begin{center}
\includegraphics*[width=7cm]{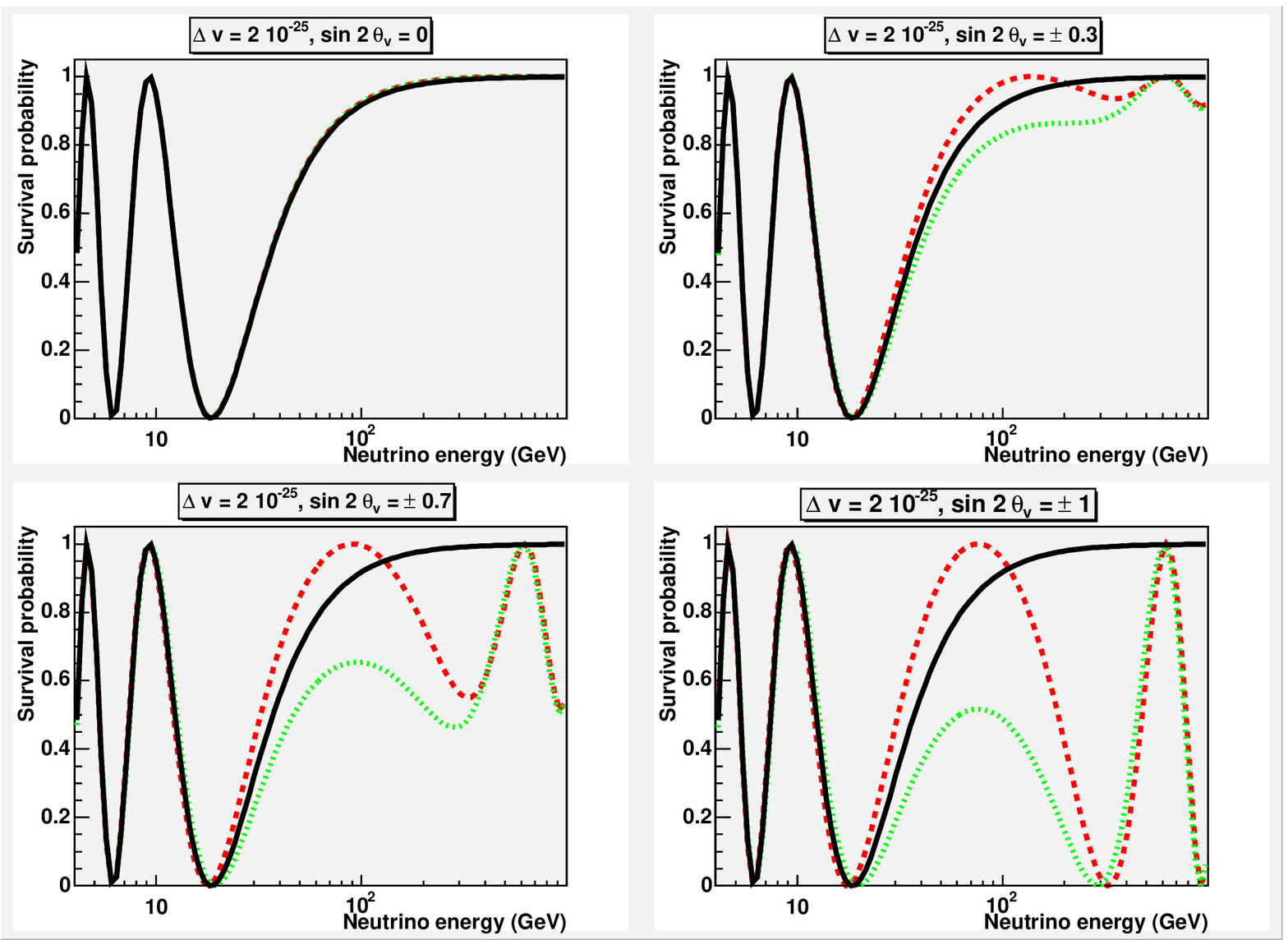}
\end{center}
\caption{Energy dependence of the $\nu_{\mu} \rightarrow \nu_{\mu}$ survival probability for 
  mass-induced oscillations alone (continuous line) 
  and mass-induced + VLI oscillations for $\Delta v = 2 \cdot 10^{-25}$ 
  and 
  sin $2\theta_v$ = $\pm$1,
  dashed lines for positive values, dotted lines for negative values.
  The neutrino pathlength was fixed at $L$ = 10000 km.} 
\label{fig:ftest}
\end{figure}

As shown in~\cite{glashow9799},
the most sensitive tests of VLI can be made by analysing the high energy tail
of atmospheric neutrinos at large pathlength values. As an example,
Fig.~\ref{fig:ftest} shows the energy dependence of the $\nu_{\mu} \rightarrow \nu_{\mu}$ survival
probability as a function of the neutrino energy, for neutrino mass-induced
oscillations alone and for both mass and VLI-induced oscillations for
$\Delta v = 2 \times 10^{-25}$ and sin$^2$ 2$\theta_v$ = $\pm$1.
Notice that for $m_{\nu} \leq 1$ eV, neutrinos with
energies larger than 100 GeV are extremely relativistic, with Lorentz
$\gamma$ factors larger than $10^{11}$.

\section{Experimental data and analysis}
\begin{figure}[htb]
\includegraphics[width=5.8cm]{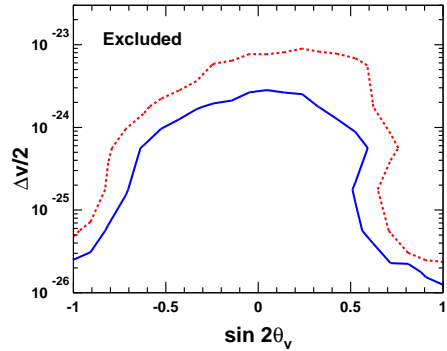}
\caption{90\% C.L. upper limits on the Lorentz invariance violation
  parameter $\Delta v$ versus sin $2\theta_v$.
  The dashed line shows the limit obtained with the selection criteria of
  Ref.~\protect\cite{mcs2}; the
  continuous line is the final result obtained with the optimized selection criteria.}
\label{fig:super}
\end{figure}

In order to analyse the MACRO data in terms of VLI, we used 
a subsample of 300 upward-throughgoing muons whose energies were estimated
via multiple Coulomb scattering in the 7 horizontal rock absorbers in the lower
apparatus~\cite{mcs1,mcs2}. The energy estimate was obtained
using the streamer tubes in drift mode, which allowed to
considerably improve the spatial resolution of the detector ($\sim$3 mm).
The overall neutrino energy resolution was of the order of 100\%,
mainly dominated by muon energy losses in the rock below the detector
(note that $\left\langle E_{\mu}\right\rangle \simeq 0.4 \left\langle  E_{\nu} \right\rangle$).
Upgoing muon neutrinos from this sample are particularly suited for
sensitive tests of VLI and VEP, since they have large energies 
($\langle E \rangle \simeq$ 50 GeV) and large zenith angles
($> 120^{\circ}$), which correspond to a median value of neutrino pathlengths
of about 10000 km. 
Following the analysis in Ref.~\cite{mcs2}, we selected a low and a high energy sample
by requiring that the reconstructed neutrino energy $E^{rec}_\nu$ should be
$E^{rec}_\nu <$ 30 GeV and $E^{rec}_\nu >$ 130 GeV.
The number of events surviving these cuts is $N_{low}$ = 49 and $N_{high}$ =
58, respectively; their median energies, estimated via Monte Carlo,
are 13 GeV and 204 GeV (assuming mass-induced oscillations).
The analysis then proceeds by fixing the neutrino mass oscillation
parameters at the values of \cite{macro-last}.
Then, we scanned the plane of the two free parameters
($\Delta v$, $\theta_v$) minimizing the $\chi^{2}$ function comprehensive of statistical and systematic uncertainties~\cite{macro-vli}.
%
%
For the Monte Carlo simulation described in~\cite{mcs2} the neutrino fluxes in input is given by \cite{newhonda}.
The largest relative difference of the extreme values of the MC expected
ratio $N_{low}/N_{high}$ is 13\%.
However, in the evaluation of the systematic error, the main sources
of uncertainties for this ratio
(namely the primary cosmic ray spectral index and neutrino cross sections)
have been separately estimated and their effects added in quadrature
(see~\cite{mcs2} for details):
in this work, we use a conservative 16\% theoretical systematic error
on the ratio $N_{low}/N_{high}$.
The experimental systematic error on the ratio was estimated to be 6\%.
%
%
The inclusion of the VLI effect does not improve the $\chi^2$ in any point
of the ($\Delta v$, $\theta_v$) plane, compared to mass-induced oscillations
stand-alone, and proper upper limits on VLI parameters were obtained.
The 90\% C.L. limits on $\Delta v$ and
$\theta_v$, computed with the Feldman and Cousins prescription~\cite{felcou},
are shown by the dashed line in Fig.~\ref{fig:super}.The energy cuts described above (the same used in Ref.~\cite{mcs2}),
were optimized for mass-induced neutrino oscillations. In order to
maximize the sensitivity of the analysis for VLI induced oscillations,
we performed a blind analysis, based only on Monte Carlo events, to
determine the energy cuts which yield the best performances. The results of this study
suggest the cuts $E^{rec}_\nu <$ 28 GeV and $E^{rec}_\nu >$ 142 GeV;
with these cuts the number of events in the real data are
$N^{\prime}_{low}$ = 44 events and $N^{\prime}_{high}$ = 35 events.
The limits obtained with this selection are shown in
Fig.~\ref{fig:super} by the continuous line.
As expected, the limits are now more stringent than for the previous choice.
%
%

An independent and complementary analysis was performed on a sample of 
events with a reconstructed neutrino energy 25 GeV $< E^{rec}_\nu <$ 75 GeV.
The number of events satisfying this condition is 106. A negative 
log-likelihood function was built event by event and then fitted to the
data. We allowed mass-induced oscillation parameters to vary inside the 
MACRO 90\% C.L. region and we left VLI parameters free in the whole 
($\Delta v$, $\theta_v$) plane.
The upper limit on the $\Delta v$ parameter resulting from this analysis
is slowly varying with $\Delta m^2$ and is of the order of $\approx 10^{-25}$.

\section{Conclusions}
We have searched for ``exotic'' contributions to standard mass-induced
atmospheric neutrino oscillations arising from a possible violation of Lorentz
invariance. Two different and complementary analyses were performed on the data, both
of them yielding compatible upper limits for the VLI contribution.
We used a subsample of MACRO  muon events for
which an energy measurement was made via multiple Coulomb scattering.
The inclusion of VLI effects does not improve the fit to
the data, and we conclude that these effects are disfavored even at
the sub-dominant level~\cite{macro-vli}. The VLI parameter bound is (at 90\% C.L.)
$|\Delta v| < 3 \times 10^{-25}$.
This result may be reinterpreted in terms of 90\% C.L. limits of
parameters connected with violation of the equivalence principle,
giving the limit $|\phi \Delta \gamma| < 1.5 \times 10^{-25}$.
The second approach exploits the information contained in a data sub-set
characterized by intermediate muon energies. It is based on the maximum likelihood
technique, and considers the mass neutrino oscillation parameters inside the 90\%
border of the global result~\cite{macro-last}.
The obtained 90\% CL upper limit on the $\Delta v$ VLI parameter
versus the assumed $\Delta m^2$ values is also around $10^{-25}$.
The limits reported in this paper are comparable to those estimated using
Super-Kamiokande and K2K data~\cite{fogli}.

\end{document}